\documentclass[11pt,a4paper,twocolumn]{IEEEtran}
\newcommand{\LineScale}{1}
\ifCLASSINFOpdf
  \usepackage[pdftex]{graphicx}
\else
  \usepackage[dvips]{graphicx}
\fi
\usepackage{color}

\usepackage{placeins}

\usepackage[cmex10]{amsmath}
\usepackage{amssymb}

\usepackage{array}
\usepackage[tight,footnotesize]{subfigure}
\usepackage{theorem}
\usepackage{enumerate}

\begin{document}
\title{Heart-induced movements in the thorax as detected by MRI}

% author names and affiliations use a multiple column layout for up to three 
% different affiliations
\author{
\IEEEauthorblockN{Lars Erik Solberg\IEEEauthorrefmark{1}},
%\IEEEauthorblockN{{\O}yvind {\AA}rdal\IEEEauthorrefmark{2}},
%\IEEEauthorblockN{Tor Berger\IEEEauthorrefmark{2}},\\
\IEEEauthorblockN{Ilangko Balasingham\IEEEauthorrefmark{1}\IEEEauthorrefmark{3}},
\IEEEauthorblockN{Erik Fosse\IEEEauthorrefmark{1}\IEEEauthorrefmark{2}},\\
\IEEEauthorblockN{Per Kristian Hol\IEEEauthorrefmark{1}\IEEEauthorrefmark{2}},\\
%\IEEEauthorblockN{Svein-Erik Hamran\IEEEauthorrefmark{2}\IEEEauthorrefmark{3}},\\
\IEEEauthorblockA{\IEEEauthorrefmark{1}Interventional Centre, Oslo University Hospital, 0424 Oslo, Norway},\\
\IEEEauthorblockA{\IEEEauthorrefmark{2}Institute of Clinical Medicine, University of Oslo, 0424 Oslo, Norway},\\
\IEEEauthorblockA{\IEEEauthorrefmark{3}Department of Electronics and Telecommunications, \\Norwegian University of Science and Technology (NTNU), 7491 Trondheim, Norway}
 \\%\IEEEauthorblockA{\IEEEauthorrefmark{2}Forsvarets forskningsinstitutt, PO Box 25, 2027 Kjeller, Norway},\\
%\IEEEauthorblockA{\IEEEauthorrefmark{3}Department of Informatics, University of Oslo, PO Box 1080 Blindern, 0316 Oslo, Norway}
}

\renewcommand{\today}{3 January 2014}


%Eldon Tyrell\IEEEauthorrefmark{4} }
%Telephone: (800) 555--1212, Fax: (888) 555--1212}


\maketitle
% Import common commands:

% LAYOUT
\newcommand{\HEADINGVSPACE}   {0.5cm}
\newcommand{\BXVSPACE}   {0.5cm}
\renewcommand{\emph}[1] {{\it\textcolor{red}{#1}}}
\newcommand{\veryemph}[1] {{\bf{\it\textcolor{red}{#1}}}}

% OLD: \newcommand{\bx}[1]      {\par \vspace{\BXVSPACE} \framebox{\parbox{\textwidth}{ #1 }}\vspace{\BXVSPACE}}
\newcommand{\bx}[1]      {\par\vspace{\BXVSPACE}\framebox{\begin{minipage}{\textwidth} #1 \end{minipage}}\par\vspace{\BXVSPACE}}
\newcommand{\eqn}[1]     {\begin{eqnarray*} #1 \end{eqnarray*}}
\newcommand{\eqnn}[1]    {\begin{eqnarray} #1 \end{eqnarray}}
\newcommand{\eqnbx}[1]   {\bx{\begin{eqnarray*} #1 \end{eqnarray*}}}
\newcommand{\eqnnbx}[1]  {\bx{\begin{eqnarray} #1 \end{eqnarray}}}
\newcommand{\eqnnmp}[2]  [0.48\textwidth]{\begin{minipage}{#1}\eqnn{#2}\end{minipage}}
\newcommand{\Mp}[2]      [0.48\textwidth]{\begin{minipage}{#1}#2\end{minipage}}
\newcommand{\items}[1]   {\begin{itemize} #1 \end{itemize}}
\newcommand{\itmb}       {\begin{itemize} \item }
\newcommand{\itmbe}[1]   {\begin{itemize} \item #1 \end{itemize}}
\newcommand{\itme}       {\end{itemize}}
\newcommand{\enums}[2] [1]{\begin{enumerate}[#1] #2 \end{enumerate}}
\newcommand{\enmbe}[2] [1]{\begin{enumerate}[#1] \item #2 \end{enumerate}}
\newcommand{\Table}[4]   {\begin{table}[h]\begin{center}\begin{tabular}{#1} #4 \end{tabular}\caption{\label{table:#2}#3}\end{center}\end{table}}
\newcommand{\SidewaysTable}[4] {\begin{sidewaystable}[h]\begin{center}\begin{tabular}{#1} #4 \end{tabular}\caption{#3}\end{center} \label{table:#2}\end{sidewaystable}}
% The following commands introduce figures. 
% Fig is the original approach: width, file, caption. 
% FigH, FigW, FigG have been added for allowing to set aspect ratios (FigG) or else 
% specifically defining height rather than width (FigH) which suggests adding one for
% specifically setting width (FigW), although this is how the original is in fact 
% working.
\newcommand{\myfigure}[3]     []{\begin{figure}[htb] \begin{center} #3 \caption{\label{figure:#2}#1}\end{center} \end{figure}}
\newcommand{\mysubfig}[3] [0.45]{\subfigure[#3]{\label{figure:#2}\includegraphics[width=#1\textwidth]{./figures/#2.eps}}}
\newcommand{\mysubfigRot}[3] [0.45]{\subfigure[#3]{\label{figure:#2}\includegraphics[height=#1\textwidth,angle=90]{./figures/#2.eps}}}
\newcommand{\Fig}[3]     {\begin{figure}[htb] \begin{center} \includegraphics[width=#1]{./figures/#2.eps} \caption{#3} \label{figure:#2} \end{center} \end{figure}}
\newcommand{\FigH}[3]    {\begin{figure}[htb] \begin{center} \includegraphics[height=#1]{./figures/#2.eps} \caption{#3} \label{figure:#2} \end{center} \end{figure}}
\newcommand{\FigW}[3]    {\begin{figure}[htb] \begin{center} \includegraphics[width=#1]{./figures/#2.eps} \caption{#3} \label{figure:#2} \end{center} \end{figure}}
\newcommand{\FigG}[3]    {\begin{figure}[htb] \begin{center} \includegraphics[#1]{./figures/#2.eps} \caption{#3} \label{figure:#2} \end{center} \end{figure}}
\newlength{\wrapfigurelength}
\newcommand{\wFig}[4] [-2cm]{\begin{wrapfigure}{l}{#2} \setlength{\wrapfigurelength}{#2}  \addtolength{\wrapfigurelength}{-0.5cm} \vspace{-1cm}\begin{center} \includegraphics[width=\wrapfigurelength]{./figures/#3.eps} \caption{#4} \label{figure:#3} \end{center} \vspace{#1} \end{wrapfigure}}
\newlength{\ImageW} 
\setlength{\ImageW}{0.75\textwidth}
\newlength{\SUBFIGUREDUALWIDTH} 
\setlength{\SUBFIGUREDUALWIDTH}{8cm}
\newlength{\SUBFIGUREDUALHEIGHT} 
\setlength{\SUBFIGUREDUALHEIGHT}{5cm}
\newcommand{\FigTwoHC}[6] [height=\SUBFIGUREDUALHEIGHT]{\begin{figure}[htb]\begin{center}%
\subfigure[#3]{\label{figure:#2}\includegraphics[width=\SUBFIGUREDUALWIDTH,#1]{./figures/#2.eps}}%
\hspace{0.5cm}
\subfigure[#5]{\label{figure:#4}\includegraphics[width=\SUBFIGUREDUALWIDTH,#1]{./figures/#4.eps}}%
\caption{#6}\end{center}\end{figure}}
\newcommand{\figureframe}[2]{\begin{figure}[htb]\begin{center} #2 \caption{#1}\end{center}\end{figure}}
\newcommand{\includesubfigurefig}[3][height=\SUBFIGUREDUALHEIGHT]{\subfigure[#3]{\label{figure:#2}\includegraphics[width=\SUBFIGUREDUALWIDTH,#1]{./figures/#2.eps}}}
\newcommand{\s}          {\quad}
\newcommand{\ls}         {\;\;}
\newcommand{\shs}        {\,}
\newcommand{\web}[1]     {http://#1}
\newcommand{\arry}[2] [ll]{\begin{array}{#1} #2 \end{array}}
\newcommand{\unit}[1]    {\left[ #1 \right]}
\newcommand{\heading}[1] {\par\vspace{\HEADINGVSPACE}\textbf{#1}\par}

% MATH EXPRESSIONS
%   - Number sets
\newcommand{\N}             {\mathbb{N}}
\newcommand{\Z}             {\mathbb{Z}}
\newcommand{\Q}             {\mathbb{Q}}
% The following is necessary in slides environments. However, it is an imperfect
% fix: this command is used for slide transitions and may cause problems. 
\ifx\R\undefined
\newcommand{\R}             {\mathbb{R}}
\else
\renewcommand{\R}           {\mathbb{R}}
\fi
\newcommand{\REAL}          {\R}
\newcommand{\C}             {\mathbb{C}}
\newcommand{\F}             {{\cal F}}% Generic field
\newcommand{\Empty}         {\emptyset}
\newcommand{\Card}[1]       {Card\left(#1\right)}
\newcommand{\intervalOO}[1] {\left \langle #1 \right \rangle}
\newcommand{\intervalCO}[1] {\left [\left. #1 \right \rangle \right.}
\newcommand{\intervalOC}[1] {\left \langle \left . #1 \right] \right .}
\newcommand{\intervalCC}[1] {\left[ #1 \right]}
\newcommand{\OneToN}[1]  [n]{\left[1,#1\right]}
\newcommand{\ZeroToN}[1] [n]{\left[0,#1\right]}
\newcommand{\Sphere}[1]     {{\bf\cal S}\left( #1 \right)}
\newcommand{\intersec}      {\cap}
\newcommand{\union}         {\cup}
\newcommand{\bintersec}     {\bigcap}
\newcommand{\bunion}        {\bigcup}
\newcommand{\interseci}[1]  {\underset{#1}{\bigcap}}
\newcommand{\unioni}[1]     {\underset{#1}{\bigcup}}

%   - operators
\newcommand{\Curl}[1] {\nabla \times #1}
\newcommand{\Div}[2]  []{\nabla_{#1} \cdot #2}
\newcommand{\Grad}[2] []{\nabla_{#1} #2}
\newcommand{\Laplace}[1] {\nabla^2 #1}
\newcommand{\vct}[1]  {{\mathbf #1}}
\newcommand{\Dyd}[1]  {\check{{\mathbf #1}}} % Dyadic
\newcommand{\dn}[3]   {\frac{\textit{d}^{#1}{#2}}{\textit{d}{#3}^{#1}}}
\newcommand{\dd}[2]   {\dn{}{#1}{#2}}
\newcommand{\dsh}[2][n]{{#2}^{(#1)}}
\newcommand{\dx}[1]   {\frac{\textit{d}{#1}}{\textit{d}x}}
\newcommand{\dy}[1]   {\frac{\textit{d}{#1}}{\textit{d}y}}
\newcommand{\dz}[1]   {\frac{\textit{d}{#1}}{\textit{d}z}}
\newcommand{\dr}[1]       {\frac{\textit{d}{#1}}{\textit{d}r}}
\newcommand{\dphi}[1]     {\frac{\textit{d}{#1}}{\textit{d}\phi}}
\newcommand{\dtheta}[1]   {\frac{\textit{d}{#1}}{\textit{d}\theta}}
\newcommand{\dv}[1]   {\frac{\textit{d}{#1}}{\textit{d}v}}
\newcommand{\dt}[1]   {\frac{\textit{d}{#1}}{\textit{d}t}}
\newcommand{\td}[1]   {\dot{#1}}
\newcommand{\tdd}[1]  {\ddot{#1}}
\newcommand{\ddt}[1]  {\frac{\textit{d}^2{#1}}{\textit{d}t^2}}
\newcommand{\ddx}[1]  {\frac{\textit{d}^2{#1}}{\textit{d}x^2}}
\newcommand{\ddy}[1]  {\frac{\textit{d}^2{#1}}{\textit{d}y^2}}
\newcommand{\ddz}[1]  {\frac{\textit{d}^2{#1}}{\textit{d}z^2}}
\newcommand{\ddr}[1]       {\frac{\textit{d}^2{#1}}{\textit{d}r^2}}
\newcommand{\ddphi}[1]     {\frac{\textit{d}^2{#1}}{\textit{d}\phi^2}}
\newcommand{\ddtheta}[1]   {\frac{\textit{d}^2{#1}}{\textit{d}\theta^2}}

\newcommand{\Jacobian}[2] {\frac{\delta(#1)}{\delta(#2)}}
\newcommand{\pd}[2]    {\frac{\delta {#1}}{\delta {#2}}}
\newcommand{\pdx}[1]   {\frac{\delta {#1}}{\delta x}}
\newcommand{\pdy}[1]   {\frac{\delta {#1}}{\delta y}}
\newcommand{\pdz}[1]   {\frac{\delta {#1}}{\delta z}}
\newcommand{\pdv}[1]   {\frac{\delta {#1}}{\delta v}}
\newcommand{\pdt}[1]   {\frac{\delta {#1}}{\delta t}}
\newcommand{\pddt}[1]  {\frac{\delta ^2{#1}}{\delta t^2}}
\newcommand{\pddx}[1]  {\frac{\delta ^2{#1}}{\delta x^2}}
\newcommand{\pddy}[1]  {\frac{\delta ^2{#1}}{\delta y^2}}
\newcommand{\pddz}[1]  {\frac{\delta ^2{#1}}{\delta z^2}}

\newcommand{\pdn}[3]   {\frac{\delta ^{(#1)} #2}{\delta {#3}^{#1}}}
\newcommand{\pdnt}[2]  {\frac{\delta ^{(#1)} #2}{\delta t^{#1}}}
\newcommand{\pdnx}[2]  {\frac{\delta ^{(#1)} #2}{\delta x^{#1}}}
\newcommand{\pdny}[2]  {\frac{\delta ^{(#1)} #2}{\delta y^{#1}}}
\newcommand{\pdnz}[2]  {\frac{\delta ^{(#1)} #2}{\delta z^{#1}}}

\newcommand{\ds}      {\vct{ds}}
\newcommand{\dl}      {\vct{dl}}
\newcommand{\Mod}     {\s mod \s}
\newcommand{\conv}    {\star}
\newcommand{\corr}    {\star\star}
\newcommand{\Fourier}[2]   []{{\cal F}_{#1}\left\{ #2 \right\}}
\newcommand{\iFourier}[2]  []{{\cal F}^{\tiny -1}_{#1}\left\{ #2 \right\}}
\newcommand{\DTFT}[2]   []{{\cal DTFT}_{#1}\left\{ #2 \right\}}
\newcommand{\iDTFT}[2]  []{{\cal DTFT}^{-1}_{#1}\left\{ #2 \right\}}
\newcommand{\FFourier}[1]  {{\cal F}_{2D}\left\{ #1 \right\}}
\newcommand{\iFFourier}[1] {{\cal F}_{2D}^{-1}\left\{ #1 \right\}}
\newcommand{\LaplaceT}[1]  {{\cal L}\left\{#1\right\}}
\newcommand{\iLaplaceT}[1] {{\cal L}^{-1}\left\{#1\right\}}
\newcommand{\ZTrans}[1]    {{\cal Z}\left\{#1\right\}}
\newcommand{\iZTrans}[1]   {{\cal Z}^{-1}\left\{#1\right\}}
\newcommand{\Hilbert}[1]  {{\cal H}\left\{ #1 \right\}}
\newcommand{\iHilbert}[1] {{\cal H}^{-1}\left\{ #1 \right\}}
\newcommand{\conj}[1] {\overline{#1}}
\newcommand{\ip}[3]   []{\left\langle #2,#3 \right\rangle_{#1}}
\newcommand{\comp}    {\circ}
\newcommand{\PseudoInv}[1] {{#1}^{\dagger}}

%   - logical operators
\newcommand{\AND}                {\,\wedge\,}
\newcommand{\ANDn}[2]          []{\,\overset{#1}{\underset{\small #2}{\bigwedge}}\,}
\newcommand{\IMPLY}              {\,\Rightarrow\,}
\newcommand{\NOT}                {\,\neg\,}
\newcommand{\OR}                 {\,\vee\,}
\newcommand{\EQUIV}              {\,\Leftrightarrow\,}
\newcommand{\st}                 {\,|\,}
\newcommand{\Cong}               {\,\equiv\,}
\newcommand{\Def}                {\,\triangleq\,}
\newcommand{\gaeq}               {\,\succsim\,}
\newcommand{\laeq}               {\,\precsim\,}

%   - Set operations:
\newcommand{\set}[1]             {\left\{ #1 \right\}}
\newcommand{\seq}[1]             {\left( #1 \right)}
\newcommand{\nuple}[2]        [n]{\left( #2 \right)_{#1}}
\newcommand{\setdiff}            {\backslash}
\newcommand{\intersection}       {\cap}
\newcommand{\family}[2]          {\left(#1_{#2}\right)_{#2}}
\newcommand{\Func}[3]            {#1:#2 \rightarrow #3}
\newcommand{\closure}[1]         {\overline{#1}}
\newcommand{\nin}                {\ni}
\newcommand{\Compl}[2]         []{\textbf{C}_#1 #2}

% The ``not in'' sign should be changed!

%   - Vector spaces notation:
\newcommand{\subspace}  {\hookrightarrow}
\newcommand{\closedsubspace}  {\overline{\hookrightarrow}}
\newcommand{\sumspaces} {\bigoplus}
\newcommand{\class}[1]  {\left[#1\right]}
\newcommand{\norm}[2]       []{\left|\left| #2 \right|\right|_{#1}}
\newcommand{\seminorm}[1]     {\mbox{semi}_{\left|\left| \cdot \right|\right|}\left(#1\right)}
\newcommand{\linfuncspace}[1] {{#1}^{\#}}
\newcommand{\Span}[1]         {Sp\left(#1\right)}
\newcommand{\ClSpan}[1]       {\overline{\Span{#1}}}
\newcommand{\Ball}[2]       []{{\cal B}_{#1}\left(#2\right)}
\newcommand{\Spectrum}[2]   []{\sigma_{#1}\left(#2\right)}
\newcommand{\OrthComp}[1]   []{{#1}^{\perp}}

% Linear operators
\newcommand{\Ker}            [1] {Ker\left(#1\right)}
\newcommand{\Image}          [1] {Im\left(#1\right)}
\newcommand{\Rank}[1]            {Rank\left(#1\right)}
\newcommand{\Dim}[1]             {Dim\left(#1\right)}
\newcommand{\Dual}[1]            {{#1}^*}
\newcommand{\Adjoint}[1]         {{#1}^*}
\newcommand{\AnnhilR}[1]         {{#1}^{\circ}}
\newcommand{\AnnhilL}[1]         {{}^{\circ}{#1}}
\newcommand{\Tpose}[1]           {{#1}^T}
\newcommand{\HTpose}[1]          {{#1}^H}
\newcommand{\Identity}[1]      []{\vct{I}_{#1}}
\newcommand{\Zero}[1]      []{\vct{0}_{#1}}

% Geometry / vector calculus:
\newcommand{\ee} {\vct{e}}
\newcommand{\ex} {\vct{e}_x}
\newcommand{\ey} {\vct{e}_y}
\newcommand{\ez} {\vct{e}_z}
\newcommand{\ephi} {\vct{e}_{\phi}}
\newcommand{\etheta} {\vct{e}_{\theta}}
\newcommand{\er} {\vct{e}_r}
\newcommand{\zero}{\vct{0}}

%   - Statistics
\newcommand{\expected}[2]        []{E_{#1}\left[ #2 \right]}
\newcommand{\bias}[2]            []{{\cal B}^{#1}\left[ #2 \right]}
\newcommand{\variance}[2]        []{VAR^{#1}\left[ #2 \right]}
\newcommand{\covariance}[2]      []{COV^{#1}\left[ #2 \right]}
\newcommand{\FisherInformation}[1] {{\cal I}\left( #1 \right)}
\newcommand{\fisherInformation}[1] {i\left( #1 \right)}
\newcommand{\RV}[1]             [X]{{\cal #1}}
\newcommand{\rv}[1]             [x]{#1}
\newcommand{\fRV}[2]      [{\cal X}]{f_{#1}\left(#2\right)}
\newcommand{\FRV}[2]      [{\cal X}]{F_{#1}\left(#2\right)}
\newcommand{\MGF}[2]      [{\cal X}]{M_{#1}\left(#2\right)}
\newcommand{\ChF}[2]      [{\cal X}]{\phi_{#1}\left(#2\right)}
\newcommand{\fRS}[2]  [{\bf\cal X}]{\fRV[#1]{#2}}
\newcommand{\FRS}[2]  [{\bf\cal X}]{\FRV[#1]{#2}}
\newcommand{\RS}[1]             [X]{{\bf \RV[#1]}}
\newcommand{\rs}[1]             [x]{{\bf \rv[#1]}}
\newcommand{\Estimator}[1] [\vct{\theta}]{\hat{#1}}
\newcommand{\Estimate}[1]  [\vct{\theta}]{\overset{*}{#1}}
\newcommand{\Prob}[2]            []{P_{#1}\left(#2\right)}
\newcommand{\normalus}             {{\cal N}\left(\mu,\sigma^2\right)}
\newcommand{\normal}[1]            {{\cal N}\left(#1\right)}
\newcommand{\cnormal}[1]            {{\cal N}_{\C}\left(#1\right)}
\newcommand{\binomial}[2]        [n]{b_{#1}\left(#2\right)}
\newcommand{\poisson}[2]         [\lambda]{p_{#1}\left(#2\right)}
\newcommand{\DistAs}             {\leadsto}
\newcommand{\Likelihood}[1]      [\vct{\theta};\RS]{L\left(#1\right)}
\newcommand{\Loglikelihood}[1]   [\vct{\theta};\RS]{l\left(#1\right)}
\newcommand{\ERisk}[1]             {{\cal R}\left(#1\right)}
\newcommand{\Risk}[1]              {{\cal R}\left(#1\right)}
\newcommand{\RiskSet}[1]           {{\cal R}}
\newcommand{\Loss}[1]              {{\cal L}\left(#1\right)}
\newcommand{\tdLikelihood}[1]    [\vct{\theta};\RS]{\td{L}\left(#1\right)}
\newcommand{\tdLoglikelihood}[1] [\vct{\theta};\RS]{\td{l}\left(#1\right)}
\newcommand{\tddLikelihood}[1]   [\vct{\theta};\RS]{\tdd{L}\left(#1\right)}
\newcommand{\tddLoglikelihood}[1][\vct{\theta};\RS]{\tdd{l}\left(#1\right)}
\newcommand{\Score}[1]           [\vct{\theta};\RS]{s\left(#1\right)}

\newcommand{\likelihood}[1]      [\vct{\theta};\rs]{L\left(#1\right)}
\newcommand{\loglikelihood}[1]   [\vct{\theta};\rs]{l\left(#1\right)}
\newcommand{\tdlikelihood}[1]    [\vct{\theta};\rs]{\td{L}\left(#1\right)}
\newcommand{\tdloglikelihood}[1] [\vct{\theta};\rs]{\td{l}\left(#1\right)}
\newcommand{\tddlikelihood}[1]   [\vct{\theta};\rs]{\tdd{L}\left(#1\right)}
\newcommand{\tddloglikelihood}[1][\vct{\theta};\rs]{\tdd{l}\left(#1\right)}
\newcommand{\score}[1]          [\vct{\theta};\rs]{s\left(#1\right)}

%   - General functions
\newcommand{\abs}[1] {\left| #1 \right|}
\newcommand{\GoesTo}[1] []{\;\overset{#1}{\rightarrow}\;}
\newcommand{\Lim}[2] {\underset{#1}{lim}\left[ #2 \right]}
\newcommand{\Limean}[2] {\underset{#1}{l.i.m}\left[ #2 \right]}
\newcommand{\Min}[2]    {\,\underset{#1}{\mbox{min}}\left\{#2\right\}}
\newcommand{\Max}[2]    {\,\underset{#1}{\mbox{max}}\left\{#2\right\}}
\newcommand{\Inf}[2]    {\,\underset{#1}{\mbox{inf}}\left\{#2\right\}}
\newcommand{\Sup}[2]    {\,\underset{#1}{\mbox{dup}}\left\{#2\right\}}
\newcommand{\Supp}[2] []{\,\underset{#1}{\mbox{dupp}}\left\{#2\right\}}
\newcommand{\LimSup}[2] {\underset{#1}{lim \; sup}\left[ #2 \right]}
\newcommand{\Arg}[2]    {\,\underset{#1}{\mbox{arg}}\left\{#2\right\}}
\newcommand{\Argmax}[2] {\,\underset{#1}{\mbox{arg max}}\left\{#2\right\}}
\newcommand{\Argmin}[2] {\,\underset{#1}{\mbox{arg min}}\left\{#2\right\}}
\newcommand{\ArgInf}[2]    {\,\underset{#1}{\mbox{arg Inf}}\left\{#2\right\}}
\newcommand{\ArgSup}[2]    {\,\underset{#1}{\mbox{arg Sup}}\left\{#2\right\}}
\newcommand{\MathFunction}[3] []{\,\mbox{#2}{#1}\left(#3\right)}
\newcommand{\MF}[2]           {{#1} \left( #2 \right) }
\newcommand{\db}[1]           {10 \cdot \Log{10}{ #1 } }
\newcommand{\Dirac}[1]        {\MF{\delta}{#1}}
\newcommand{\Heaviside}[1]    {\MF{u}{#1}}
\newcommand{\Sin}[2]        []{\MathFunction{sin$^{#1}$}{#2}}
\newcommand{\Cos}[2]        []{\MathFunction{cos$^{#1}$}{#2}}
\newcommand{\Sec}[2]        []{\MathFunction{sec$^{#1}$}{#2}}
\newcommand{\Csc}[2]        []{\MathFunction{csc$^{#1}$}{#2}}
\newcommand{\Tan}[2]        []{\MathFunction{tan$^{#1}$}{#2}}
\newcommand{\Cot}[2]        []{\MathFunction{cot$^{#1}$}{#2}}
\newcommand{\Sinh}[2]        []{\MathFunction{sinh$^{#1}$}{#2}}
\newcommand{\Cosh}[2]        []{\MathFunction{cosh$^{#1}$}{#2}}
\newcommand{\Tanh}[2]        []{\MathFunction{tanh$^{#1}$}{#2}}
\newcommand{\Arcsin}[2]       []{\MathFunction{arcsin$^{#1}$}{#2}}
\newcommand{\Arccos}[2]       []{\MathFunction{arccos$^{#1}$}{#2}}
\newcommand{\Arctan}[2]       []{\MathFunction{arctan$^{#1}$}{#2}}
\newcommand{\Log}[3]          []{\MathFunction[^{#1}_{#2}]{log}{#3}}%{log_{#1}\left(#2\right)}
\newcommand{\Lnz}[2]          []{\MathFunction[^{#1}]{Ln}{#2}}
\newcommand{\Ln}[2]           []{\MathFunction[^{#1}]{ln}{#2}}
\newcommand{\Imag}[1]         {\MathFunction{im}{#1}}
\newcommand{\Real}[1]         {\MathFunction{re}{#1}}
\newcommand{\Sgn}[1]          {\MathFunction{sgn}{#1}}
\newcommand{\Rect}[2]         []{\MathFunction{rect$^{#1}$}{#2}}
\newcommand{\Sinc}[2]         []{\MathFunction{sinc$^{#1}$}{#2}}
\newcommand{\Tri}[2]          []{\MathFunction{tri$^{#1}$}{#2}}
\newcommand{\Pdf}[2]          {\mbox{pdf}_{#1}\left(#2\right)}
\newcommand{\Cdf}[2]          {\mbox{cdf}_{#1}\left(#2\right)}
\newcommand{\Zarg}[1]         {\mbox{Arg}\left(#1\right)}
\newcommand{\Ceil}[1]         {\left\lceil  #1 \right\rceil}
\newcommand{\Floor}[1]        {\lfloor #1 \rfloor}
\newcommand{\Trace}[1]        {\MathFunction{tr}{#1}}
\newcommand{\Exp}[1]          {\MathFunction{exp}{#1}}
\newcommand{\diag}[1]         {\MathFunction{diag}{#1}}
\newcommand{\GammaF}[1]       {\MathFunction{$\Gamma$}{#1}}

%   - Other mathmatical constructions
\newcommand{\Matrix}[2]      {\left[\begin{array}{#1} #2 \end{array}\right]}
\newcommand{\Determinant}[2] {\left|\begin{array}{#1} #2 \end{array}\right|}
\newcommand{\jacobian}[2]    {\frac{\delta\left(#1\right)}{\delta\left(#2\right)}}
\newcommand{\Degrees}        {{^\circ}}
\newcommand{\Order}[1]       {{\cal O}\left(#1\right)}
\newcommand{\Poynting}       {{\cal P}}
\newcommand{\Comb}[2]        {\left( {#2 \atop #1} \right)}

%   - Hyperlinks and references (need to include package hyperref in frame.tex
\newcommand{\Link}[3]      [blue]{\hyperlink{#2}{\textcolor{#1}{#3}}} % May refer to any \label, in addition \Target may be used
\newcommand{\Target}[3]    [blue]{\hypertarget{#2}{\textcolor{#1}{#3}}}

\newcommand{\thref}  [1]{theorem \ref{th:#1}}
\newcommand{\thsref} [1]{\ref{th:#1}}
\renewcommand{\eqref}  [1]{(\ref{eq:#1})}
\newcommand{\eqsref}   [1]{(\ref{eq:#1})}
\newcommand{\figref}   [1]{fig. \ref{figure:#1}}
\newcommand{\figsref}  [1]{figs. \ref{figure:#1}}
\newcommand{\Figref}   [1]{Fig. \ref{figure:#1}}
\newcommand{\Figsref}  [1]{Figs. \ref{figure:#1}}
\newcommand{\tablref}  [1]{tabl. \ref{table:#1}}
\newcommand{\tablsref} [1]{tabls. \ref{table:#1}}
\newcommand{\Tablref}  [1]{Tabl. \ref{table:#1}}
\newcommand{\Tablsref} [1]{Tabls. \ref{table:#1}}
\newcommand{\secref}   [1]{section \ref{section:#1}}
\newcommand{\secsref}  [1]{sections \ref{section:#1}}
\newcommand{\Secref}   [1]{Section \ref{section:#1}}
\newcommand{\Secsref}  [1]{Sections \ref{section:#1}}

%   - Common texts
\newcommand{\first}{$1^{st}$}
\newcommand{\second}{$2^{nd}$}
\newcommand{\third}{$3^{rd}$}
\newcommand{\vOne}{\vct{1}}
\newcommand{\vZero}{\vct{0}}

%   - Specific symbols
\newcommand{\SFA} {\overset{SFA}{\approx}}%{{SFA \atop \leftrightarrow}} % Stationnary Phase Approximation
\newcommand{\TEQ}[1] []{\overset{#1}{\leftrightarrow}} % Euiqvalence by transformation
\newcommand{\Ohm} {\Omega}

\newcommand{\review}[2][]{\textcolor{blue}{{\it #1}}\textcolor{red}{#2}}

% Add project-specific commands:
\newcommand{\AFig}[4] {
\begin{figure}[htb] 
\begin{center} 
\includegraphics[width=#2]{./figures/#3.eps}
\caption{#4}
\label{figure:#1}
\end{center}
\end{figure}}

\renewcommand{\Fig}[5] [0 cm]{
\begin{figure}[t] 
\begin{center} 
\hspace{#1}
\includegraphics[width=#3]{./figures/#4.eps}
\caption{#5}
\label{figure:#2}
\end{center}
\end{figure}}

\newcommand{\Question}[1] {{\textcolor{blue}{#1}}} % Empty

\newcommand{\NA}[1]{-}

\newlength{\CW}
\setlength{\CW}{\LineScale\columnwidth} % Compensate for single or double line spacing.
\setlength{\ImageW}{0.9\CW}
\renewcommand{\Table}[4] {\begin{table}[h]\begin{center}\caption{\label{table:#2}#3}\begin{tabular}{#1} #4 \end{tabular}\end{center}\end{table}}
\renewcommand{\mysubfig}[3] [0.45\CW]{\subfigure[#3]{\label{figure:#2}\includegraphics[width=#1]{./figures/#2.eps}}}

\renewcommand{\Exp}[1] {e^{#1}}
\renewcommand{\Imag}[1]         {\MathFunction{Im}{#1}}
\renewcommand{\Real}[1]         {\MathFunction{Re}{#1}}
\renewcommand{\Sgn}[1]          {\MathFunction{sgn}{#1}}

\renewcommand{\iFourier}[2] []{FT^{\tiny -1}_{#1}\left\{ #2 \right \}}
\renewcommand{\Fourier}[2]  []{FT_{#1}\left\{ #2 \right \}}

% Remove figure contents (for light-weight printing):
% \renewcommand{\mysubfig}[3] [0.0]{}
% \renewcommand{\myfigure}[3] {}
% \renewcommand{\Fig}[3] {}

\renewcommand{\emph}[1] {{\it #1}}
\newcommand{\B} {{\cal B}}

\renewcommand{\unit}[1] {\shs#1}

\renewcommand{\figref}   [1]{Fig. \ref{figure:#1}}
\renewcommand{\figsref}  [1]{Figs. \ref{figure:#1}}

%********************************************************************************
\begin{abstract}
%\boldmath
In order to provide information for the use of radar in diagnostics a qualitative map of movements in the thorax has been obtained.

This map was based on magnetic resonance image sequences of a human thorax during suspended respiration. The movements were measured using two distinct techniques. Segmentation provided measures of aorta dilatation and displacements, and image edge detection indicated other movements. 

The largest heart movements were found in the anterior and left regions of the heart with in-plane displacements on the order of 1 cm and which caused lung vessels displacements on the order of 2-3mm especially on the left side due to the heart ventricular. 

Mechanical coupling between the heart and aorta caused aorta displacements and shape distortions. Despite this coupling, aorta dilatations most likely reflected blood pressure variations.

\end{abstract}

%\tableofcontents



% \fi
% \IEEEpeerreviewmaketitle

%********************************************************************************
\section{Introduction}
This article addresses the movements of tissues in a human thorax as detected by magnetic resonance images (MRI) of a single, healthy individual. Motion in the thorax is primarily due to respiration of the lungs and to the beating heart. 

The motion induced by respiration is extensively studied in the literature. Several studies have quantified the precision of motion compensation in imaging processes \cite{Siebenthal2007,Hinkle2009,McLeish2008,Paganelli2013}, although a quantized map of movements is generally not presented. F. Odille and colleagues have, however, presented a map in a sagital plane for one study \cite{Odille2008}. Larger movements due to respiration is found along the superior-inferior direction with anterior-posterior being the second most significant axis \cite{Brandner2006} \cite{Weber2009b}, \cite{Wang1995}, \cite{Langen2001}.

Dilatations of the aorta are found to be due to the blood flow through the aorta, and the aortic displacements are both a consequence of respiration and the heart beats \cite{Biesdorf2011}, \cite{Rengier2012}, \cite{Weber2009}. Motion induced specifically by the heart activity, e.g. while suspending respiration, is reviewed in \cite{Scott2009} with an emphasis on improving cardiac imaging, however, no map is provided.

While investigating radar-based techniques for estimating aortic blood pressure based on characteristics of the aorta geometry a map in the axial plane of heart-induced motions was needed, as was a more detailed description of the aortic behavior. We have not been successful in finding an adequate map in the literature. This article hence provides a simple map of organ boundary movements in general and aortic dilatations and displacements in particular. During all acquisitions, respiration was suspended.

\section{Methods}
\label{section:methods}
Two sets of  magnetic resonance images (MRI) were recorded: one set (I) for constructing a 3D volume representation; one set of axial planes (II) for analyzing tissue movements through the heart cycle. A three Tesla MR system (Achieva, Phillips Healthcare, Best, The Netherlands) was used. In this presentation, a right-handed coordinate system was used with 'x' pointing from left to right (lateral), 'y' pointing from posterior towards anterior (depth) and 'z' pointing from lower-back towards the neck (longitudinal) (\figref{Overview}).

In set I, a voxel element spanned a volume of $0.89 \times  0.89 \times 1 \unit{\mbox{mm}^3}$ using a transverse, Turbo Field Echo procedure. For image acquisition the respiration was not suspended, but synchronization with respiration was implemented at close to minimal inhalation.

For set II, a balanced, Turbo Field Echo procedure was used where the heart cycle was divided into 30 phases, axially oriented, for each of the longitudinal positions. Each slice had a 2D slice pixel size of  $0.85 \times  0.85 \unit{\mbox{mm}^2}$, at 6.6 mm slice spacing from just below the heart to slightly below the neck. For each acquisition, respiration was suspended; however, only a partial inhalation was done in order to allow for good ECG quality.  

In both sets, MRI images were recorded in synchronization with the ECG signal and the sets were both T1-weighted.

The images of set II were further segmented with relation to the aorta in order to have precise ``position'' and ``size'' data. The segmentation was based on a manually initialized, elliptic seed with automatic growth in a transformed neighborhood while attempting to maintain region homogeneity. The ``position'' of the aorta was given through the centroid of the segmented pixels in each image. The ``size'' was defined as the radius of the circle with equal area as the segmented ellipsis: $\hat{r} = \sqrt{ab}$, where $a,b$ are the minor and major axes. This procedure was not considered successful in the lower most axial slices: in these slices, strong artifacts in the MR images due to blood flow precluded reliable segmentation.

For other structures, segmentation was considered too difficult and a different approach was adopted in order to illustrate the tissue displacements: this was based on the extraction of intensity gradients (edges) and which were assumed to be tissue boundaries. First the edges ($E$) of the compressed MRI images ($J$) in set II ($S$) were calculated; these are binary images. These were then accumulated ($M$) over the set of time instants.
\eqn{\forall I_{z,t}\in S, \s J_{z,t} &=& \sqrt{I_{z,t}} \in \R^{N_x \times N_y} \\
      E_{z,t} &=& edge(J_{z,t};\theta,\sigma) \in \set{0,1}^{N_x \times N_y} \\
      M_{z} &=& \Sigma_{t=1}^{N_t} E_{z,t} \in  \set{0,\cdots,N_t}^{N_x \times N_y}}
where $t$ is here an index into the heart cycle (with $N_t$ instants) and the variables are of size $N_x \times N_y$. The edge detection was the Canny edge detector, with thresholds set by $\theta$ and size set by $\sigma$.

Ideally, movements created sets of similar edges in a neighborhood with extent depending on the scale of movement, and static edges would all be identified at the same point $(x,y)$ and $M_{z}(x,y)=N_t$. In practice, the accumulated edges are not unambiguous indications of movements (see results in \figref{Other-Movements}):
\enmbe[i]{\label{es:spurr}``thin'', gray lines represented spurious detections (e.g. point(3)),
\item\label{es:ill_def}  a dark gray, narrow ($\approx$2 pixels) line may represent ill-defined boundaries (examples along the lung-thoracic cage boundary),
\item \label{es:plane} intensity changes and boundaries which passed in/out of the plane may have created numerous, ``parallel'' lines which nonetheless did not represent the movement of a single boundary (e.g. point(6)).
}
Regarding \ref{es:spurr} and \ref{es:ill_def}, edges with poor quality were not always detected, or not always at the same position. 

The accumulated edges image was chosen to represent movements because it allowed the sequence of edges to be displayed in a single two-dimensional image. A movie representation avoids some of the aforementioned problems and was the basis of adding qualitative indications of displacements in the form of blue arrows to these images.

Remark, displacements assessed in a two dimensional plane can only provide reliable estimations of the component in the plane, not orthogonal to the plane (z): actual boundary movements were necessarily less.

\section{Results}
\label{section:results}
The variation in aortic radius was virtually equal for all slices with very similar profiles (to within the noise of the measurements): this profile was saw-tooth shaped (\Figref{Aorta_size_var}). Neither the dicrotic notch nor shoulder was clearly apparent in these measurements - this may have been due to lack of segmentation accuracy.

Through the heart cycle, the aorta moved along a grossly diagonal path in the sagital planes as demonstrated by the aorta centroids and particularly so in the upper torso (correlation between $x$ and $y$ in \figref{Aorta_centroid_var}). This was due to the aorta being pushed against the spine which constrained the movements. The amplitude of the movement in $y$ is roughly 1.5 mm. Lower down the torso, a more complex pattern emerges. 

The difference in mechanical coupling with the heart as a function of z-level does not appear as a difference in profile of the aorta size.

\Fig[-0.5 cm]{Overview}{1.1\ImageW}{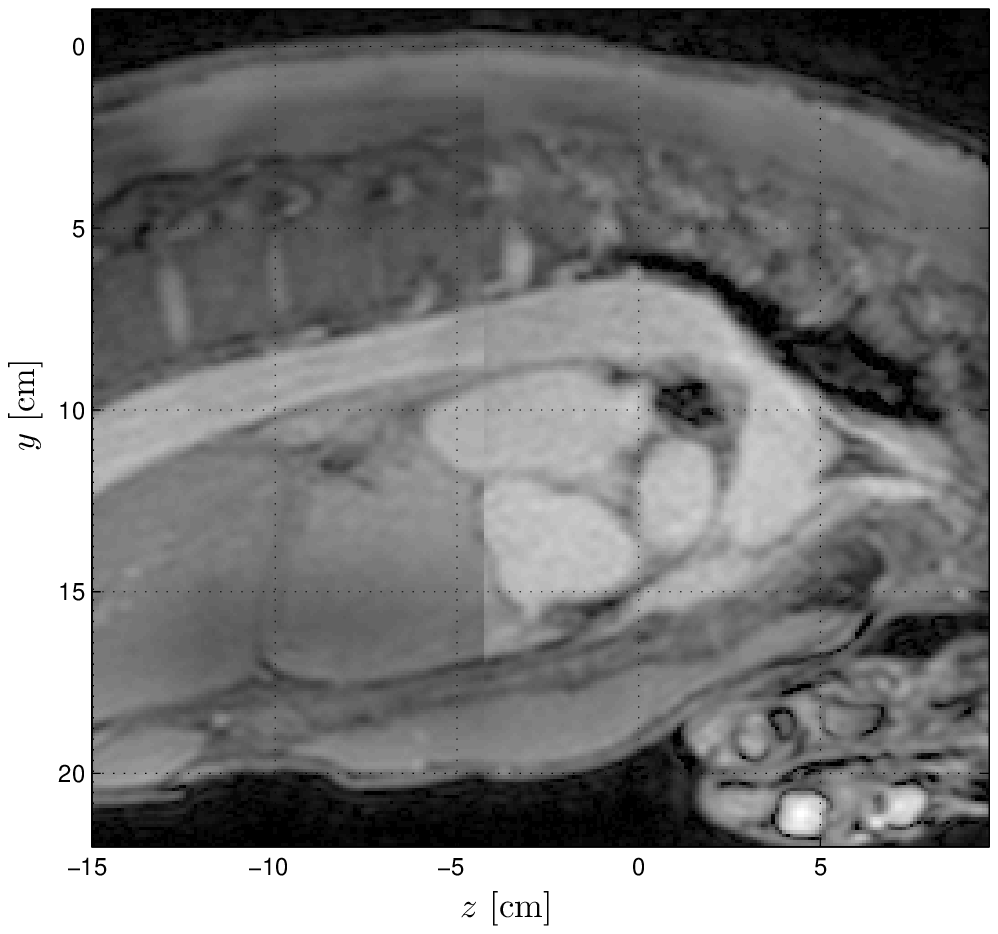}{A sagittal section through the subject and which is centered on the aorta ($x = -1.0cm$) visible as a long, tube beneath the spine. The heart can be seen as an oval, oblique structure in the center with several visible heart chambers and just below the aorta. The positive $x$ axis points out of the image.}

\Fig[-1 cm]{Aorta_centroid_var}{1.2\ImageW}{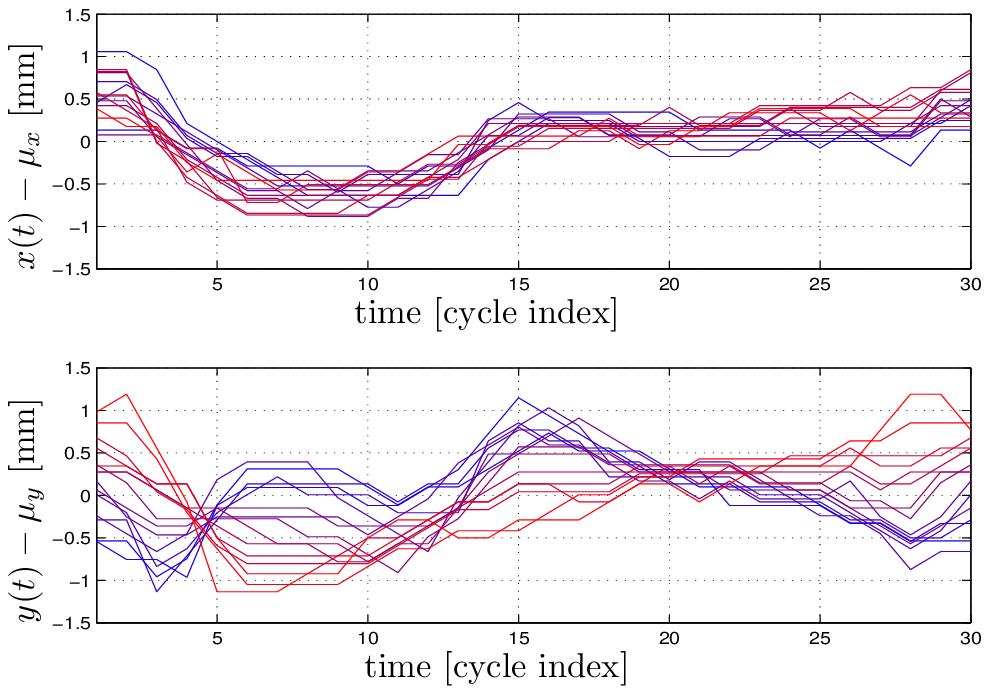}{Aorta centroid (top: $x$, bottom: $y$) displacements as a function of time through a heart cycle and subtracted mean values ($\mu_x$, $\mu_y$) to ease comparison. Upper z levels red; lower levels blue: $z \in \intervalCC{-6.9,2.3}$ cm.}

\Fig[-1 cm]{Aorta_size_var}{1.2\ImageW}{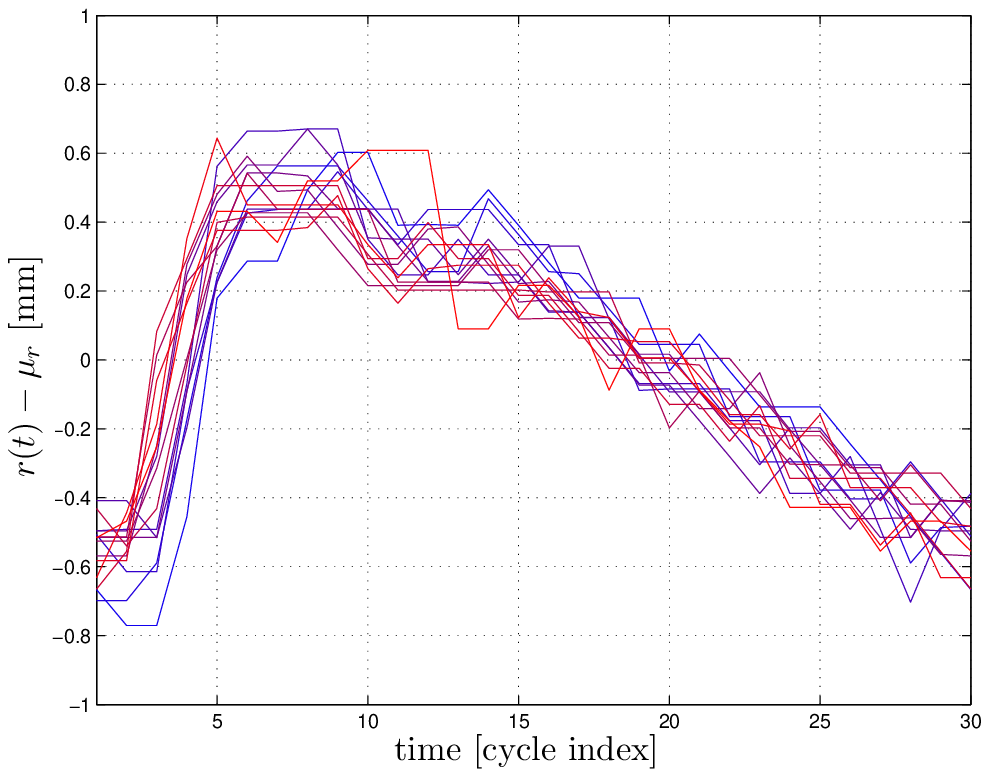}{Aorta radius dilatations as a function of time through a heart cycle and subtracted mean values to ease comparison. Upper z levels red; lower levels blue: $z \in \intervalCC{-6.9,2.3}$ cm. The radius profiles indicate temporal shifts while descending the aorta: red traces appear to precede blue traces.}

\Fig{aortic_blood_pressure}{\ImageW}{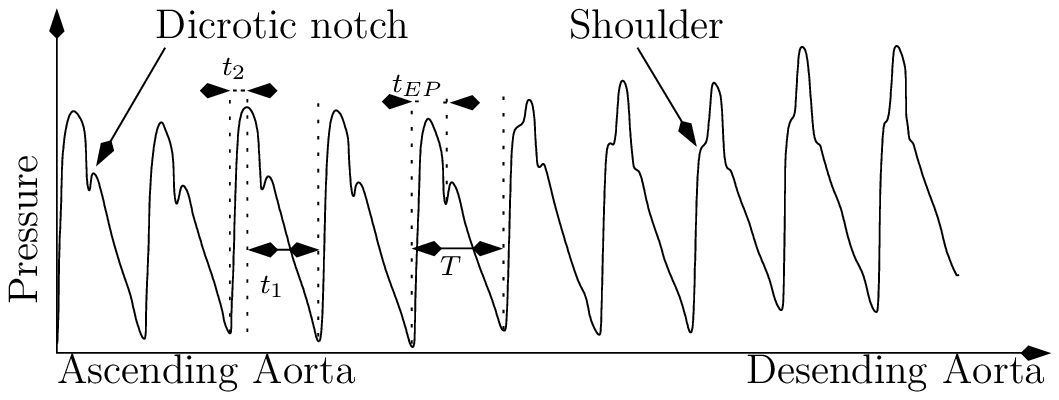}{The series of blood pressure cycles is reproduced from Murakami \cite{Murakami2005} and was measured in a human while gradually removing the catheter from close to the aortic valve. Typical features of the aortic pressure wave is shown: the dicrotic notch at the closing of the aortic valve, the ``shoulder'' associated with the reflected wave from further down the aorta. Finally, the form of the pressure pulse changes as the reflection arrives at different time within the cycle. The actual blood pulse's shape depends on the individual.}

The accumulated edges $M_z$ shows large movements of the heart with induced movements in neighboring tissues, particularly large blood vessels in the lungs. These movements were strongest to the left, probably due to left ventricular activity. Edges appeared also to be detected between the compartments of the heart, and several of these internal interfaces show large displacements. The right side of the heart showed comparatively smaller displacements, therefore with smaller induced movements in neighboring tissues.

Very little traces of movements were detected between the lungs and the thoracic cage.

\Fig{Other-Movements}{\ImageW}{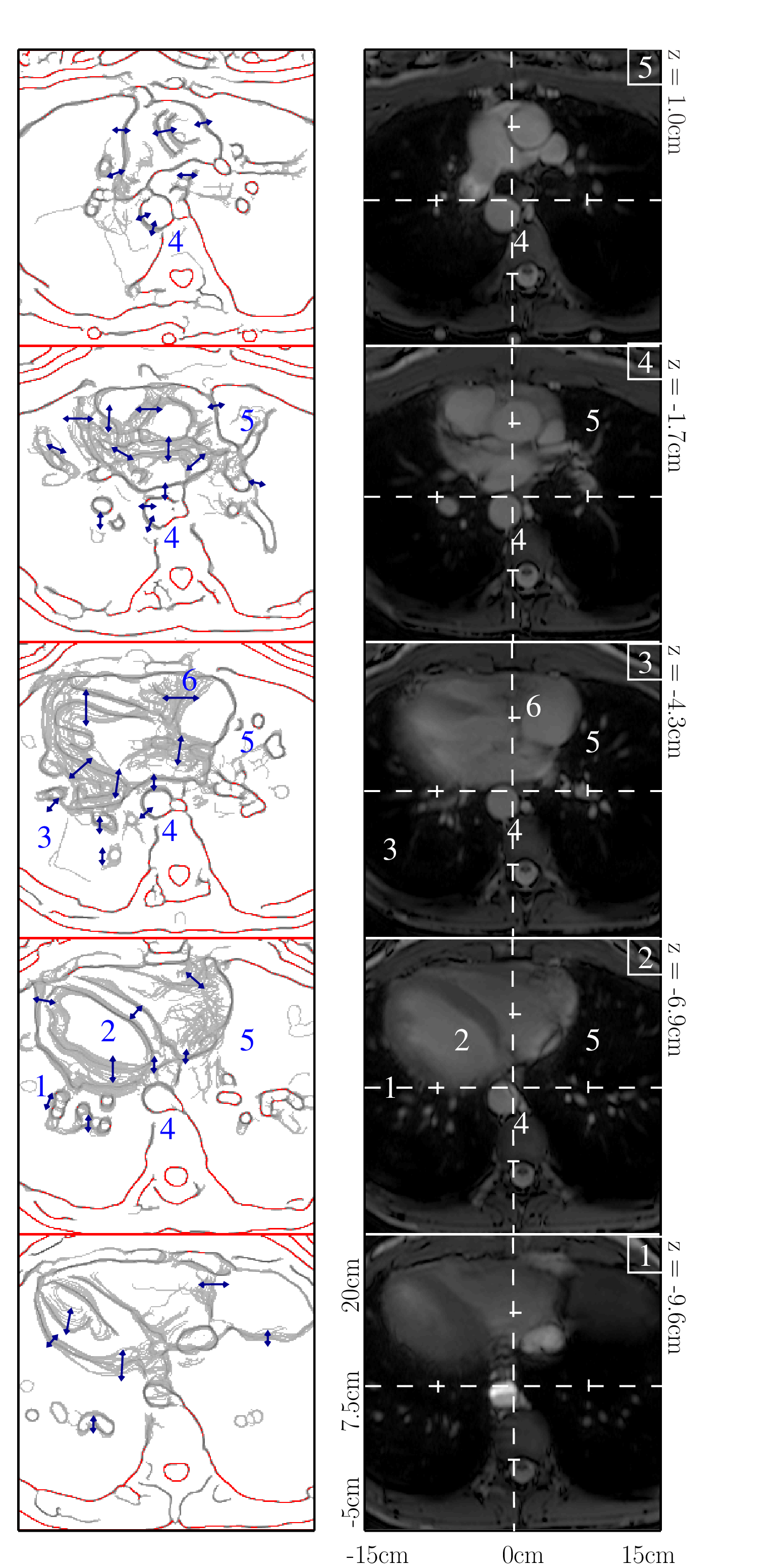}{To left, accumulated edges $M_z$; to right average intensities ${\expected[t]{J_{z,t}}}$. All images span the same $xy$ region; in images 1 through 5 $z$ was in the range $\intervalCC{-9.6,1.0}$ cm. In the edge images, the red lines represent boundaries that essentially did not move (more than 24 of 30 images in the same point). Sets of neighboring, similar curves are likely due to tissue boundaries that were moving. The blue arrows are qualitative measures of movements based on edge ``movies'' ($E_{z,t}$ as time sequences). In image 1, blood flow in the aorta created artifacts.}

\section{Discussion}
The largest movements were observed in the heart and dominated by the left ventricular. Outside the heart, the tissue boundaries displaced by the heart were generally those between lungs and arteries. This coupling has two possible causes: the blood flow through the arterial tree and the pressure propagation through intermediate tissues. Given that the blood flow through the arterial tree should act similarly in the left and right sides while apparent movements were strongest to the left, the dominant effect was interpreted to be due to the propagation of the left ventricular movements.

Regarding the scale of in-plane displacements of the heart walls, large movements were generally on the order of 0.5 cm to 1 cm with the largest exceeding 1 cm, perhaps as much as 2 cm (point (6) in image 3, \figref{Other-Movements}). The displacements were typically divided into 4 phases: long even compression, long retraction, still, short retraction. Outside the heart, the displacement scales were significantly less and the stronger, observable displacements appeared to be on the order of 2-3 mm.

Although the scale of the lung arteries' movements are greater than that of the aorta, the former constitutes a chaotic tree structure with relatively thin branches as opposed to the aorta which is larger and runs almost straight down the back. 

Close to the aorta, heart wall displacements appeared to have lesser amplitude. This movement was coupled with the aorta dilatations and in some instances (e.g. images 3, 4 in \figref{Other-Movements}) resulted in complex deformations and displacements. This was probably due to an interaction between internal blood pressure pulses and the mechanical coupling.

The segmentation of the aorta resulted in measures of both radius and centroid. The segmentation followed a semi-automatic method and although the aorta was a fairly easy organ to extract in MRIs, artifacts are known to affect the images in particular due to strong blood flow or any large movement in general. This was also the reason why the lower-most z-levels were excluded from \figsref{Aorta_centroid_var} and \ref{figure:Aorta_size_var}. For those slices included in these figures, the continuity of the traces across slices and relatively low noise along a trace both indicated an acceptable precision of the segmentation.

As mentioned in the previous section, the aorta centroid followed a grossly diagonal path, especially at upper z levels. In \figref{Other-Movements}, the aorta is located between the heart wall and the spine - at least in the upper z levels. Heart pulsations exhorted pressure on the aorta which in turn was constrained by the spine: this may explain the diagonal movement. The tendency to adjust to the heart pressure is seen in \figref{Other-Movements}: the accumulated edges indicates stronger movements of the aorta wall along this SW-NE diagonal. These movements were partially correlated with the aorta size: compare the relative $x$ movements with aorta radius. The reason for this correlation was likely the small lag between the blood pressure pulse wave through the aorta and the mechanical coupling with the nearby heart walls.

In \cite{Stefanidis1995} the radius dilatations of the aorta were measured using a precise and invasive method based on pressure and diameter sensors introduced through catheters. It concluded that typical radius peak-to-peak amplitudes for a normal population is 1.09 $\pm$ 0.22 mm. In \cite{Rengier2012}, similar results were obtained at several locations along the aorta. The scale of dilatations in \figref{Aorta_size_var} are of the same size. 

The linear relationship between relative changes of blood pressure and cross sectional area was studied in \cite{Sugawara2000}: it provided a link between the radius changes and blood pressure variations. A similar study \cite{Olsen1972} also shows the similarity of pressure and diameter using trans-oesophagal ultrasound. The form of the aorta radius dilatations was similar to common central arterial blood pressure curves (\figref{aortic_blood_pressure}).

Furthermore, the lack of correlation between aorta size and $y$ position as functions of the $z$ level pointed out in the results section proved at least that size and position are partially decorrelated. Given the lack of visible influence, if not independent, the influence of $y$ on size was small. Finally, we saw no reason to assume that the influence of $x$ on aorta size should be stronger then that of $y$.

Therefore, we concluded that the relationship between size and internal pressure was likely to be valid in spite of the pressure exerted on the aorta from the heart.

The analysis in this article has been based on sequences of MRI images. Given that the in-plane resolution of the images were approximately 1 mm, the scale of movements detectable in the accumulated edges images was on the order of this resolution. Given that the aorta was segmented in each MRI acquisition and that the measures of size and centroid were based on the entire segmented regions, their accuracy was significantly higher than the 1 mm bound.

The observations of this article relied only on a single individual in prone position. The posture of a given individual is likely to affect certain aspects - in particular how the aorta is squeezed between the heart and the spine. However, the general map of displacements is likely to remain valid.

\section{Conclusion}
In this article, a map of heart-induced tissue displacements in the thorax is presented. Due to limited resolution, movements of scale less than 1 mm were not detected.

The largest heart movements were found in the anterior and left regions in the heart; the scale of these in-plane displacements were on the order of 1 cm and caused artery-lung boundary displacements on the order of 2-3mm. These induced displacements were far stronger in the left than right lungs due to the proximity to the heart left ventricular.

Mechanical coupling between the heart and aorta caused aorta centroid displacements with axial variations. This coupling could be seen to affect the aorta wall in a complex manner. However, due to lacking correlation with displacement, similarity with typical blood pressure curves and expected range of diameter variations, aorta radius variations was concluded to be dominated by internal pressure variations.

\section{Acknowledgments}
This work is part of the MELODY project, which is funded by the Research Council of Norway under the contract number 187857/S10. 
% We would like to express gratitude to the Department of Radiology and Nuclear Medicine, Oslo University Hospital in Oslo, Norway for providing the MR images.
The Department of Radiology and Nuclear Medicine, Oslo University Hospital in Oslo, Norway provided the MR images.

%********************************************************************************




\bibliographystyle{IEEEtran}
\bibliography{IEEEabrv,./references}

\end{document}